 \documentclass[doublespacing]{elsart}

\usepackage{natbib}

\usepackage{graphicx}

\usepackage{amssymb}

 \usepackage{lineno}

\begin{document}

\begin{frontmatter}



\title{Calculation of the enrichment of the giant planets envelopes during the ``late heavy bombardment''}

\author[alexis]{A. Matter\corauthref{cor}}
\corauth[cor]{Corresponding author.}
\ead{Alexis.Matter@obs-nice.fr}
\author[tristan]{T. Guillot}
\ead{Tristan.Guillot@obs-nice.fr}
\author[alessandro]{A. Morbidelli}
\ead{Alessandro.Morbidelli@obs-nice.fr}
\address[alexis]{Laboratoire Fizeau, CNRS UMR 6203, Observatoire de la C\^ote d'Azur, BP 4229, 06304 Nice Cedex 4, France.  Phone: +33 (0)4 92 00 30 68, fax: +33 (0)4 92 00 31 38}
\address[tristan]{Laboratoire Cassiop\'ee, CNRS UMR 6202, Observatoire de la C\^ote d'Azur, BP 4229, 06304 Nice Cedex 4, France. Phone: +33 (0)4 92 00 30 47, fax: +33 (0)4 92 00 31 21}
\address[alessandro]{Laboratoire Cassiop\'ee, CNRS UMR 6202, Observatoire de la C\^ote d'Azur, BP 4229, 06304 Nice Cedex 4, France. Phone: +33 (0)4 92 00 31 26, fax: +33 (0)4 92 00 31 21}

\begin{abstract}
The giant planets of our solar system possess envelopes consisting
mainly of hydrogen and helium but are also significantly enriched in
heavier elements relatively to our Sun. In order to better constrain
how these heavy elements have been delivered, we quantify the amount
accreted during the so-called "late heavy bombardment", at a time when
planets were fully formed and planetesimals could not sink deep into
the planets. On the basis of the "Nice model", we obtain accreted
masses (in terrestrial units) equal to $0.15\pm0.04 \rm\,M_\oplus$ for
Jupiter, and $0.08 \pm 0.01 \rm\,M_\oplus$ for Saturn. For the two other giant planets, the results are found to depend mostly on whether they switched position during the instability phase. For Uranus, the accreted mass is $0.051 \pm 0.003 \rm\,M_\oplus$ with an inversion and $0.030 \pm 0.001 \rm\,M_\oplus$ without an inversion. Neptune accretes $0.048 \pm 0.015 \rm\,M_\oplus$ in models in which it is initially closer to the Sun than Uranus, and $0.066 \pm 0.006 \rm\,M_\oplus$ otherwise. With well-mixed
envelopes, this corresponds to an increase in the enrichment over the solar value of $0.033 \pm 0.001$ and $0.074 \pm 0.007$ for Jupiter and Saturn, respectively. For the two other planets, we find the enrichments to be $2.1 \pm 1.4$ (w/ inversion) or $1.2 \pm 0.7$ (w/o inversion) for Uranus, and $2.0 \pm 1.2$ (w/ inversion) or $2.7 \pm 1.6$ (w/o inversion) for Neptune. This is clearly insufficient to
explain the inferred enrichments of $\sim 4$ for Jupiter, $\sim 7$ for
Saturn and $\sim 45$ for Uranus and Neptune.
\end{abstract}


\end{frontmatter}
\linenumbers
\section{Introduction}

The four giant planets of our solar system have hydrogen and helium envelopes which are enriched in heavy elements with respect to the
solar composition. In Jupiter, for which precise measurements from the Galileo probe are available, C, N, S, Ar, Kr, Xe are all found to be enriched compared to the solar value by factors 2 to 4~\citep{1999Natur.402..269O,2004Icar..171..153W} (Assuming solar abundances based on the compilation by \citet{2003ApJ...591.1220L}). In Saturn, the C/H ratio is found to be 7.4 $\pm$ 1.7 times solar \citep{2005Sci...307.1247F}. In Uranus and Neptune it is
approximately $45 \pm 20$ times solar \citep{GuillotGautier2006} (corresponding to about $30$ times solar with the old solar abundances). Interior models fitting the measured gravitational fields constrain enrichments to be between 1.5 and 8 for Jupiter and between 1.5 and 7 times the solar value for Saturn \citep{2004ApJ...609.1170S}. For Uranus and Neptune, the envelopes are not massive enough (1 to 4 Earth masses) for interior models to provide global constraints on their compositions.\\ 
Enriching giant planets in heavy elements is not straightforward. \citet{2000ASPC..219..475G} have shown that once the planets have their final masses, the ability of Jupiter to eject planetesimals severely limits the fraction that can be accreted by any planet in the system. The explanations put forward then generally imply an early enrichment mechanism:\\
\begin{itemize}
\item \citet{2005A&A...434..343A} show that migrating protoplanets can have
  access to a relatively large reservoir of planetesimals and accrete
  them in an early phase before they have reached their final masses
  and started their contraction. This requires the elements to
  be mixed upward efficiently, which is energetically possible, and
  may even lead to an erosion of Jupiter's central core
  \citep{2004jpsm.book...35G}. 
\item The forming giant planets may accrete a gas that has been
  enriched in heavy elements through the photoevaporation of the
  protoplanetary disk's atmosphere, mainly made of hydrogen and helium
  \citep{2006MNRAS.367L..47G}. This could explain the budget in noble gases
  seen in Jupiter's atmosphere but is not sufficient to explain the enrichment in
  elements such as C, N, O  because small grains are prevented from reaching the planet due to the formation of a dust-free gap \citep[eg.][]{2007A&A...462..355P}.
 The photoevaporation model requires that the giant planets form late in the evolution of disks, which appears consistent with modern scenarios of planet formation (see \citet{2004ApJ...604..388I};  \citet{2008ApJ...686.1292I}). It also implies that a significant amount of solids are retained in the disk up to these late stages, as plausible from simulations of disk evolution (e.g. \citet{2007ApJ...671.2091G}).
\end{itemize}
It has been recently suggested that the Solar System underwent a major change of structure during the phase called `Late Heavy Bombardment' (LHB) \citep{2005Natur.435..459T, 2005Natur.435..466G}. This phase, which occurred $\sim 650$~My after planet formation, was characterized by a spike in the cratering history of the terrestrial planets.\\ 
The model that describes these structural changes, often called the `Nice model' because it was developed in the city of Nice, reproduces most of the current orbital characteristics of both planets and small bodies. This model provides relatively tight constraints on both the location of the planets  and on the position and mass of the planetesimal disk at the time of the disappearance of the proto-planetary nebula. Thus, it is interesting to study the amount of mass accreted by the planets at the time of the LHB, in the framework of this model.\\
The article is organized as follows: we first describe the orbital evolution model at the base of the calculation. Physical radii of the
giant planets at the time of the late heavy bombardment are also discussed. We then present results, both in terms of a global enrichment and in the unlikely case of an imperfect mixing of the giant planets envelopes. 
\section{The `Nice' model of the LHB}
\subsection{General description}
The Nice model postulates that the ratio of the orbital periods of Saturn and Jupiter was initially slightly less than 2, so that the
planets were close to their mutual 1:2 mean motion resonance (MMR); Uranus and Neptune were supposedly orbiting the Sun a few AUs beyond
the gas giants, and a massive planetesimal disk was extending from 15.5~AU, that is about 1.5~AU beyond the last planet, up to 30--35~AU. \\
As a consequence of the interaction of the planets with the planetesimal disk, the
giant planets suffered orbital migration, which slowly increased
their orbital separation. As shown in \citet{2005Natur.435..466G} N-body
simulations, after a long quiescent phase (with a duration varying
from 300~My to 1~Gy, depending on the exact initial conditions),
Jupiter and Saturn were forced to cross their mutual 1:2 MMR. This
event excited their orbital eccentricities to values similar to those
presently observed. \\ 
The acquisition of eccentricity by both gas
giants destabilized Uranus and Neptune. Their orbits became very
eccentric, so that they penetrated deep into the planetesimal
disk. Thus, the planetesimal disk was dispersed, and the interaction
between planets and planetesimals finally parked all four planets
on orbits with separations, eccentricities and inclinations similar to
what we currently observe. \\
This model has a long list of successes. It
explains the current orbital architecture of the giant planets
\citep{2005Natur.435..459T}. It also explains the origin and the
properties of the LHB. In the Nice model, the LHB is triggered by the
dispersion of the planetesimal disk; both the timing, the duration and
the intensity of the LHB deduced from Lunar constraints are well
reproduced by the model \citep{2005Natur.435..466G}.\\
Furthermore, the Nice model also explains the capture of planetesimals around the
Lagrangian points of Jupiter, with a total mass and orbital
distribution consistent with the observed Jupiter Trojans \citep{2005Natur.435..462M}. More recently, it has been shown to provide a framework for understanding the capture and orbital distribution of the
irregular satellites of Saturn, Uranus and Neptune, but not of Jupiter \citep{2007AJ....133.1962N}, except in the case of an encounter between Jupiter and Neptune which is a rare but not impossible event. The main properties of the Kuiper belt (the relic of the
primitive trans-planetary planetesimal disk) have also been explained
in the context of the Nice model (\citet{2008Icar..196..258L}; see \citet{2008ssbn.book..275M}, for a review).
\subsection{The dynamical simulations}
In this work, we use 5 of the numerical simulations performed in
\citep{2005Natur.435..466G}. The main simulation is the one
illustrated in the figures of the Gomes et al. paper. The 1:2 MMR
crossing between Jupiter and Saturn occurs after 880~My, relatively
close to the observed timing of the LHB (650~My). When the instability
occurs, the disk of planetesimals still contained 24 of its initial 35
Earth masses ($M_\oplus$).\\ 
During the evolution that followed the
resonance crossing, Uranus and Neptune switched position. Thus,
according to this simulation, the planet that ended up at $\sim 30$~AU
(Neptune) had to form closer to the Sun than the planet that reached a
final orbit at $\sim 20$~AU (Uranus).\\ 
However, because the planets
evolutions are chaotic during the instability phase, different outcomes
can be possible. Thus Gomes et al. performed 4 additional simulations
with initial conditions taken from the state of the system in the main
simulation just before the 1:2 resonance crossing, with slight
changes in the planets' velocities. Two of these `cloned'
simulations again showed a switch in positions between Uranus and
Neptune, but the two others did not. That is, in these two cases the
planet that terminated its evolution at 30~AU also started the
furthest from the Sun. \\
Here we use these 5 simulations (the main one
and its 4 `clones') to evaluate the amount of solid material accreted
by the planets and how it could vary depending on the specific
evolutions of the ice giants. Notice that, whereas the main simulation
spans 1.2~Gy (and therefore continues for 320~My after the 1:2 MMR
crossing), the cloned simulations cover only a time-span of
approximately 20~My, and were stopped when the planets reached
well separated, relatively stable orbits. 
\subsection{Probability of impact and accretion of planetesimals}
From these dynamical simulations, several steps are necessary to 
estimate the amount of mass accreted by the planets.\\
For each simulation at each output time (every 1~My) we have the orbital
  elements of the planets and of all the planetesimals in the system. We are aware that this 
time interval may be a bit too long to resolve the evolution of the 
system during the transient phases that immediately follow the onset of 
the planetary instability. On the other hand, when the instability 
occurs, most of the disk is still located beyond the orbits of the 
planet, so that the bombardment rate is not very high. Thus, we believe 
that this coarse time sampling is enough for our purposes.\\
  First we look for planetesimals that are in a mean motion
  resonance with a planet. The resonances taken into account are the
  1:1, 1:2, 2:3, 2:1, 3:2. When computing the collision probability with a
  planet, the objects in resonance with that planet will not be taken
  into account (but they will be considered for the collision
  probability with the other planets). The rationale for this is that
  the resonant objects, even if planet-crosser, cannot collide with
  the planet, because they are phase-protected by the resonant
  configuration, as in the case of Pluto.\\  
The width of a resonance is proportionnal to $\sqrt{\frac{{\rm M_{\rm planet}}}{{\rm M_{\rm sun}}}}*{\rm a}$ where ${\rm M_{planet}}$ is the mass of the considered planet and ${\rm a}$ is the semi-major axis of the precise resonant orbit. Hence we take an approximative relative width of $\frac{\Delta a}{a}=1\%$ on the semi-major axis to define the area where a planet and a given planetesimal are considered to be in resonance. 
Then we select all the non-resonant particles that cross the orbit of a planet. The intrinsic collision probability $P_i$ of each of these particles with the planet is computed using the method detailed in \citet{1967JGR....72.2429W}, implemented in a code developed by \citet{1992A&A...253..604F} and kindly provided to us.
Once $P_i$ is known for each particle ($i=1,\ldots,N$), the mass accreted by the planet during the time-step $\Delta t$ is simply :
\begin{equation}
\nonumber
M_{acc}=\sum_{i=1}^{N} P_i R^2_{\rm planet} M_{i} \Delta t f_{\rm grav}
\end{equation}
where M$_{i}$ is the mass of the planetesimal (with M$_i=0.00349$ Earth mass), and $f_{grav}$ is
the gravitational focusing factor. The latter is equal to :
\begin{equation}
\nonumber
f_{grav}=1+\frac{V_{\rm lib}^2}{V_{\rm rel}^2}\ 
\end{equation}
where $V_{\rm rel}$ is the relative velocity between the planet and the
planetesimal, and $V_{\rm lib}$ is the escape velocity
from the planet. Finally, the total mass accreted by a planet during the
  full dynamical evolution is simply the sum of M$_{acc}$ over all
  time-steps taken in the simulation.
\section{Results: mass accreted by each giant planets}
Fig.1 shows the cumulated mass captured by Jupiter, Saturn, Uranus
and Neptune as a function of time in the case of
the main simulation. The abrupt increase at 882 Ma is due
to the triggering of the LHB when Saturn crosses the 1:2
resonance with Jupiter.  It is interesting to notice that this
short phase accounts for about two third of the mass acquired by
the planets during their full evolution.\\
Qualitatively, the more massive is the planet, the larger is
the  massed accreted from the planetesimal disk. This is because
larger planets have larger gravitational cross
sections.\\ 
 Uranus and Neptune have comparable masses, and therefore which
  planets accretes more mass depends on their orbital histories.  In
  the model shown in Fig.1, Uranus first accretes planetesimals at a
  larger rate than Neptune because Uranus is initially the furthest
  planet in the system and the closest to the planetesimal
  disk. However, when the two planets exchange position and Neptune
  is scattered into the planetesimal disk, it accretes
  many more planetesimals and eventually exceeds Uranus in terms
  of total accreted mass.
As previously described, the evolution of the system is very chaotic  
and Fig.1 only represents one of the possible outcomes. In order to  
assess the variability of the solutions, we also present in Fig.2 the  
evolution of the 4 "cloned" simulations focused around the critical  
period of the 1:2 MMR crossing. We note them 'simu {\itshape a}', 'simu {\itshape b}', 'simu {\itshape c}', and 'simu {\itshape d}'. These simulations were started at 868  
Myr, just before the 1:2 MMR crossing ($\approx 880$ Myr), and stopped at 893,  
897, 875 and 899 Myr respectively, as soon as the planets reached well  
separated and relatively stable orbits.\\ 
Fig.2 shows that the variability of the accreted masses during that period amounts to up to  
a factor 2 for all planets except Neptune, for which the variability  
is a factor 4. The added uncertainty on the results due to the 1 Myr  
timestep appears small in comparison, as shown by the regularity of  
the curves.\\
In order to obtain the evolution of the mass accreted by 4 giant  
planets during the entire 1.2 Gyr period and to assess the effect of  
the position switch between Uranus and Neptune on the final accreted  
mass, we proceed as follows:\\ 
we simply assume that the planets  
accreted the same amount of mass as in the main simulation over the  
first 880 Myr and over the time ranging from the end of each cloned  
simulation up to 1,200 Myr. In the cases in which Uranus and Neptune  
do not switch positions, we consider that Uranus accreted before the  
LHB the same mass accreted by "Neptune" (the 4th planet) in the main  
simulation, and inversely for Neptune. The results are shown in Fig.  
3. We can see that the results for simulations are very similar to the result for the main simulation. However in
the simulation {\itshape a}, Neptune eventually accretes less mass than
Uranus. Conversely, the results for the simulations {\itshape b} and {\itshape d} are
 qualitatively different. Neptune is initially the closest planet 
to the disk
and hence accretes much more planetesimals than Uranus also
before the LHB. This remains the case during/after the LHB, since Neptune is scattered into the disk 
and acquires even more planetesimals compared to Uranus.\\
Table 1 summarizes the total masses accreted by the
planets, and compares them 
to the masses of heavy elements in their hydrogen-helium envelopes estimated from interior models fitting the giant planets gravitational moments \citep[see][]{2005AREPS..33..493G}.  
As before, for Uranus and Neptune, we separate the cases in which
these planets exchange their positions from the cases in which they do
not.\\ 
In the first case, Uranus accretes an amount of
planetesimals comparable to Neptune's, whereas when the
order of the ice giants is not switched, Neptune
accretes twice more planetesimals than Uranus. In all cases, the
masses accreted are significantly smaller than the masses of the envelopes.\\ 
For Jupiter and Saturn, the mass accreted is
much lower ($\approx 10^{-3}$ times smaller), whereas
this ratio can increase to $\sim 7\times10^{-2}$ for Uranus and 
Neptune. Therefore in the framework of the 'Nice' model, the LHB has a stronger impact in terms of heavy elements supply relatively to the envelope mass, in the case of the two latter planets, Uranus and Neptune.    
\section{Calculation of the envelope enrichments}  
\subsection{Fully mixed case}
Once we calculated the mass that each planet accreted during this
period, it is straightforward to infer the corresponding change in
composition. We thus calculate the increase of the atmosphere's enrichment $\Delta\cal{E}$,
defined as the amount of heavy elements for a given mass of atmosphere
compared to that same value in the Sun. More specifically, the global
enrichment increase of a giant planet envelope of mass $M_{\rm envelope}$
accreting a mass of planetesimals $M_{\rm accreted}$ (assuming that
planetesimals don't reach the core) is:
\begin{equation}
\nonumber
{\Delta\cal E_{\rm LHB}}=\frac{M_{\rm accreted}}{M_{\rm envelope}\times Z_\odot}
\end{equation} 
where $Z_\odot$ is the mass fraction of heavy elements in the
Sun. Following \cite{2005EAS....17...21G}, we use in mean $Z_\odot=0.015$. 
This global enrichment is also the enrichment of the atmosphere,
provided the envelope is well-mixed, a reasonnable assumption given
the fact that these planets should be mostly convective \citep[see e.g.][]{2005AREPS..33..493G}.
These values of enrichment are calculated by taking the mean of the accreted masses of the table 1, and the uncertainty on the envelope mass is taken into account.
Table 2 shows that this yields relatively small enrichments: the contribution of this late veneer of
planetesimals accounts for only about $1\%$ of the total enrichments of Jupiter and Saturn, and up to $10\%$ in the case of Uranus and
Neptune, owing to their smaller envelopes.
\subsection{Incomplete mixing case}
Mixing in the envelopes of giant planets is expected to be fast
compared to the evolution timescales, and rather complete because
these planets are expected to be fully convective
\citep{2005AREPS..33..493G}. 
We want to test the possibility, however unlikely, that mixing was not
complete, and that the observed atmospheric enrichments were indeed
caused by these late impacts of planetesimals.\\
The values of enrichment, in the hypothesis of an incomplete mixing of the envelope, depends on two elements : the extent of mixing of heavy elements in the envelope, but also the penetration depth of planetesimals in the envelope as a function of their size distribution at the time of the LHB.\\
First let us evaluate to what extent the mixing should occur in order to retrieve the observed enrichments ($\cal{E}_{\rm C/H}$ in Table 2). Following Eq.3, we evaluate the mass of envelope over which planetesimals should penetrate and be mixed to explain the observations :
\begin{equation}
{\rm M}_{\rm mixed}=\frac{{\rm M}_{\rm accreted}}{\cal{E}_{\rm C/H}\times {Z_\odot}}
\end{equation}
Using hydrostatic balance, and assuming a constant adiabatic gradient of 0.3, a pressure of 1 bar at the top of the atmosphere and a gravitational acceleration constant and equal to its value at the top of the atmosphere, we calculate the corresponding penetration depth, h$_{\rm mixed}$, and pressure, P$_{\rm mixed}$, at this depth .
We now have to consider in addition the penetration depth of planetesimals in the envelope as a function of their size distribution at the time of the LHB. The reason is that, even if we could define the extents of penetration and mixing giving the observed enrichments, an important fraction of planetesimals penetrating more deeply in the envelope would anyway cause a heavy elements supply over a larger extent, implying atmospheric enrichments lower than those observed.\\
We thus have to determine the mass of the envelope shell enriched by a planetesimal of a given size. The main assumption here is to consider that during its entry into the atmosphere, the planetesimal desintegrates and mixes completely with the atmosphere after crossing a mass of gas column equal to its own mass. Thanks to a parallel plane approximation of the atmosphere, the mass of the atmospheric shell thus enriched can be inferred from the ratio between the planetary area $a_{\rm planet}$ and the planetesimal section $a_{\rm pl}$ multiplied by the mass of the considered planetesimal $M_{pl}$ : 
\begin{equation}
\nonumber
M_{\rm enriched\:shell}=M_{\rm pl}\frac{a_{\rm planet}}{a_{\rm pl}}
\end{equation}
We now define s$_{\rm mixed}$, the critical planetesimal radius for which $M_{\rm enriched\:shell}=M_{\rm mixed}$. Following Equations 5 and 6 and considering ice spherical planetesimals with a mass density noted $\rho$ :
\begin{equation}
s_{\rm mixed}=\frac{3\times M_{\rm mixed}}{4\times a_{\rm planet}\times\rho}
\end{equation}
  All the planetesimals larger than s$_{\rm mixed}$ will penetrate more deeply than the extent of mixing and will enrich a larger part of the envelope, yielding the enrichments lower than observed.\\
We now evaluate the mass fraction of planetesimals with sizes larger than s$_{\rm mixed}$.
For that we use a bi-modal size distribution inspired from the observations of Trans-Neptunian Bodies \citep{2004AJ....128.1364B}, and used successfully by \citet{2007Icar..188..468C} to explain the number of comets in the scattered disk and the Oort cloud :
\begin{equation}
\nonumber
\frac{dN}{ds}=f_{small}\times s^{-3.5} \qquad s<s_0
\end{equation}
\begin{equation}
\nonumber
\frac{dN}{ds}=f_{big}\times s^{-4.5} \qquad s>s_0
\end{equation}
with $50km <s_0< 100km$ the turnover radius, and f$_{\rm small}$ and
f$_{\rm big}$ the normalisation factors which depends on the value of
s$_0$ and the total mass of the planetesimals disk \citep{2005Natur.435..466G}. For each planet, the mass fraction is calculated with s$_0$=50 and 100 km in order to have a good range of values around the estimated one which is approximately 70~km according to \citet{2008AJ....136...83F}.\\
Table 3 summarizes the results obtained in the case of an incomplete mixing.
Compared to the whole envelope mass, the masses of layer enriched by this incomplete mixing are of the order of 1\% for Jupiter and Saturn, and between 5 and 10\% for Uranus and Neptune.    
As previoulsy mentioned, these values and those of the related quantities are a priori unrealistic because of the globally convective structure of the giant planets.
Moreover, according to the results of mass fraction in Table 3, we see that the large planetesimals with a size bigger than $s_{\rm mixed}$ are comparable and even predominant in terms of mass compared to the small ones, especially for Uranus and Neptune.\\ 
Therefore even if we assume an incomplete mixing giving the observed enrichments, the important supply of heavy elements by the large planetesimals at layers deeper than h$_{\rm mixed}$ will imply anyway lower enrichments than those observed.\\
In summary, it appears that the observed enrichments cannot be explained in the context of the Late Heavy Bombardment even by using the hypothesis of an incomplete mixing. 
\section{Conclusion}
In this work, we evaluated the extent to which the late heavy bombardment could explain the observed enrichments of giant planets.\\ We calculated the
mass accreted by each planet during this period thanks to several dynamical simulations of the LHB within the so-called "Nice" model. The accreted masses were found to be much smaller than those of the envelopes of each giant planet. In the realistic hypothesis of a global mixing in these envelopes, we found the enrichments over the solar value to be approximately two orders of magnitude smaller than the observations for Jupiter and Saturn and one order of magnitude smaller than the observations for Uranus and Neptune.\\
We then tested the possibility of an incomplete mixing in the giant planets envelopes to account for the observed enrichments. With a size distribution of planetesimals inferred from observations of trans-neptunian bodies, we found that the enrichments were always at least a factor of 2 lower than observed. Given the efficient convection expected in the deep atmosphere, we expect however the mixing to be complete. \\   
Therefore we conclude that the enriched atmospheres of the giant planets do not result from the Nice model of the LHB and probably from any model describing the LHB. In fact \citet{2000ASPC..219..475G}'s calculations showed that the mass needed to explain Jupiter's and Saturn's enrichments would be certainly much too large, in any late heavy bombardment model. Earlier events should thus be invoked in the explanation of the enriched atmopspheres of giant planets. On the other hand the enrichment process during the LHB may not be completely negligeable when considering fine measurements of the compositions of giant planets \citep[eg.][]{2008ExA...tmp...34M}. When present it may also have a role in enriching the envelopes of close-in extrasolar giant planets because of their radiative structure.\\
\\ 

We acknowledge support from the Programme National de Plan\'etologie. We thank one of the referees, Brett Gladman, for comments that improved the article.

\bibliography{LHB}

Caption figure 1 :
Mass accreted (in Earth mass units) by Jupiter (plain), Saturn (dashed), Uranus (dash-dotted) and Neptune (dotted)
respectively, as a function of time (in years). The simulation
corresponds to the main simulation described in the text, in
which Uranus and Neptune switch their relative positions.

caption figure 2 :
Additional mass accreted (in Earth mass units) during the time range of the 4 'cloned' simulations. Figures a and c (left) correspond to the cases in
which Uranus and Neptune exchange position at the time of the
LHB. Figures b and d (right) show the result of simulations in
which the four planets preserve their initial order. These 'cloned' simulations start 868 Myr after the beginning of the planets migration and are stopped once giant planets acquired well separated and stable orbits.

Caption figure 3 :
Mass accreted(in Earth mass units) for the 4 `cloned' simulations during the whole time scale of the 'Nice' model. In each of these four panels, the period before 868 Myr and after 875-899 Myr (depending on the simulation) is assumed to be identical to the main simulation. Figures a and c (left) correspond to the cases in
which Uranus and Neptune exchange position at the time of the
LHB. Figures b and d (right) show the result of simulations in
which the four planets preserve their initial order.

Caption table 1 :
Planetesimal masses accreted by the giant planets after the disappearance of
the protosolar gaseous disk. 

Caption table 2 :
Enrichment increase (see equation 3) calculated from the different
simulations of the model. The values of the observed $\epsilon_{\rm C/H}$ are derived from \citet{2005AREPS..33..493G}.

Caption table 3 :
Mass of enriched layer, extent of mixing h$_{\rm mixed}$, pressure level at the bottom of the mixing area P$_{\rm {mixed}}$, and planetesimal radii s$_{\rm mixed}$, which would match the observed enrichments. The last column is the mass percentage corresponding to the planetesimals whose the radius is larger than s$_{\rm mixed}$, the range being due to the two limiting values of s$_0$ used.

\begin{figure}[!h]
   \centering
   \caption{}
   \includegraphics[height=85mm,width=130mm,angle=0]{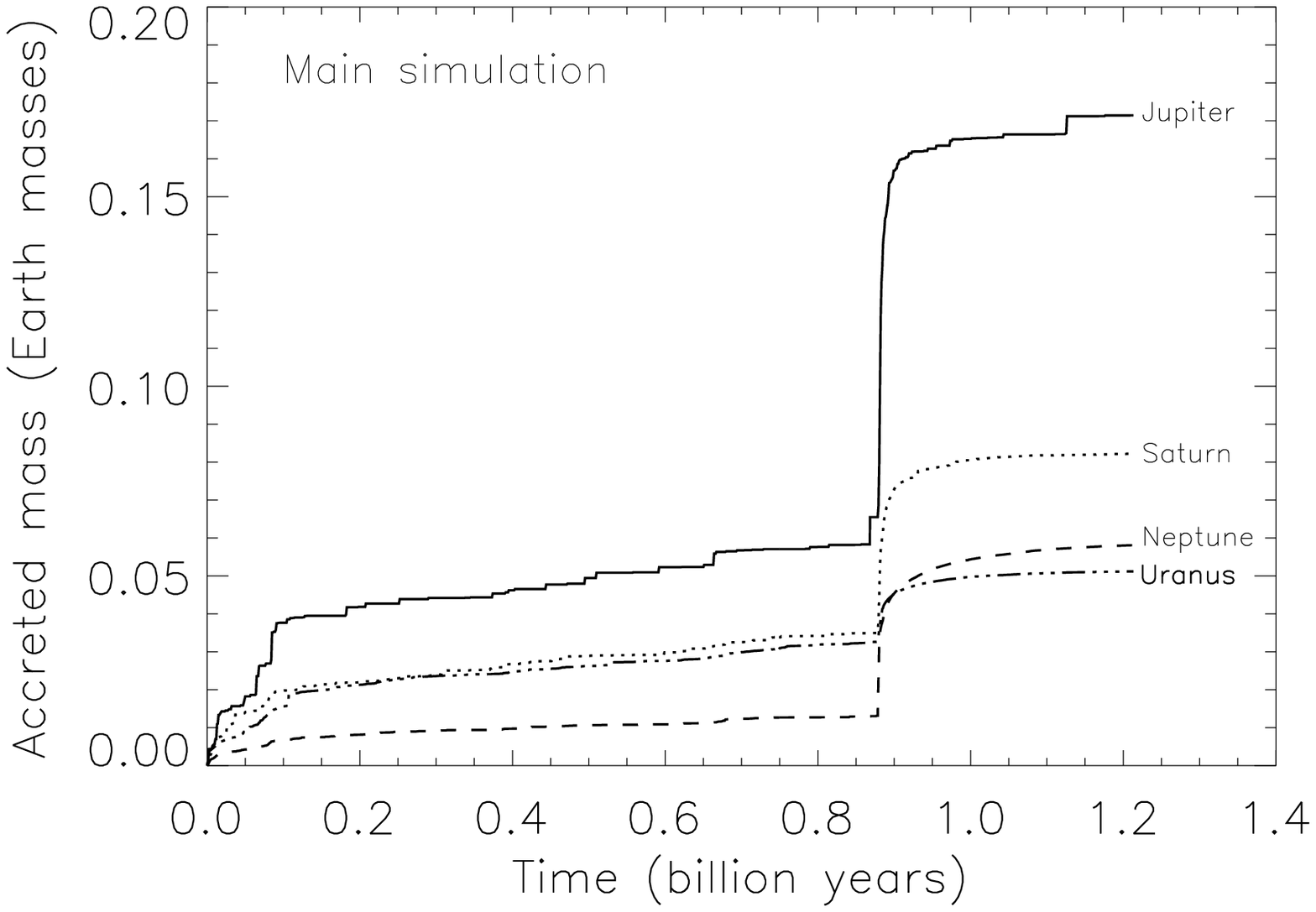}
   \label{fig:Figure}
\end{figure}

\begin{figure}[!h]
\caption{}
\includegraphics[height=57mm,width=70mm,angle=0]{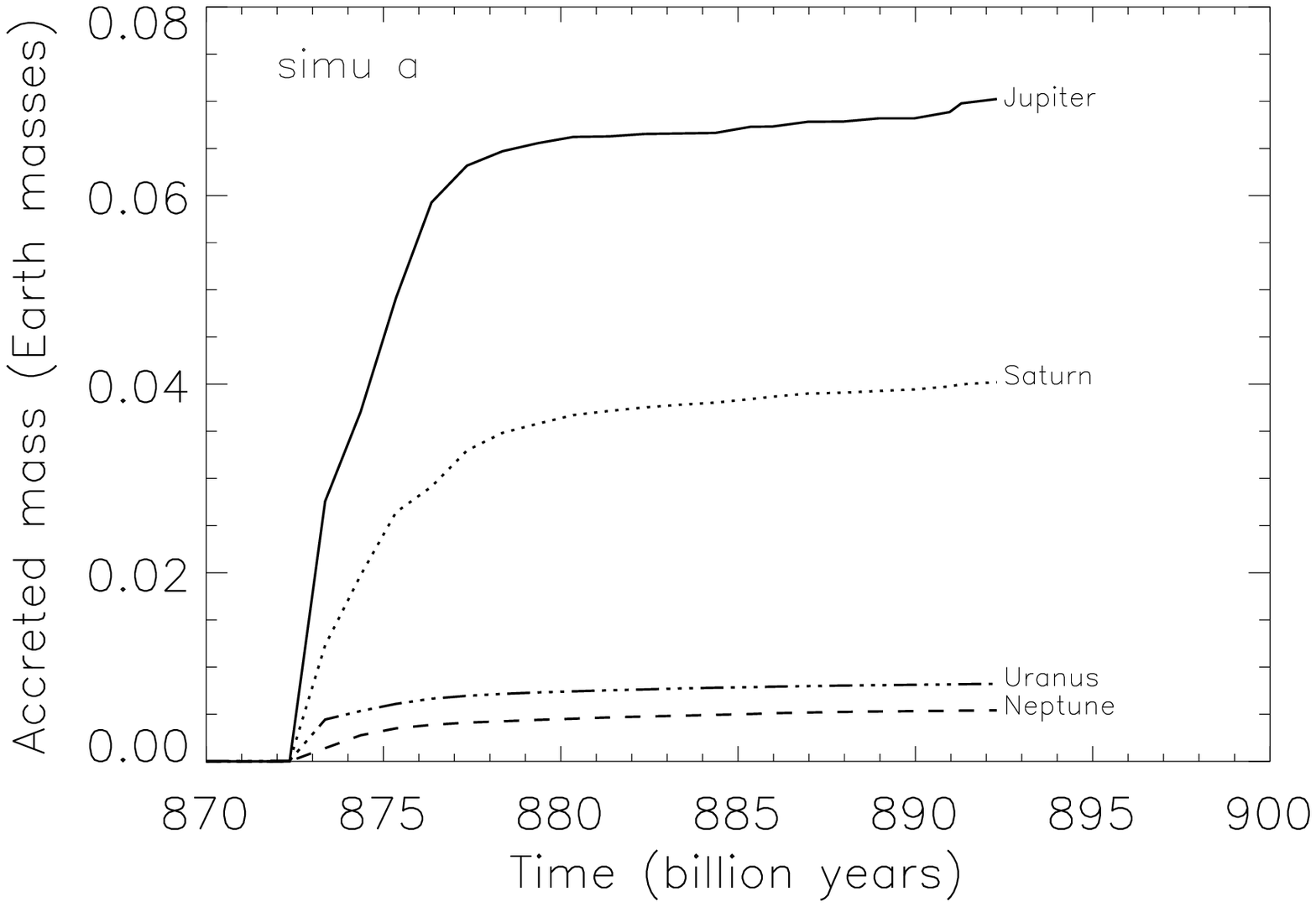}\hfill
\includegraphics[height=57mm,width=70mm,angle=0]{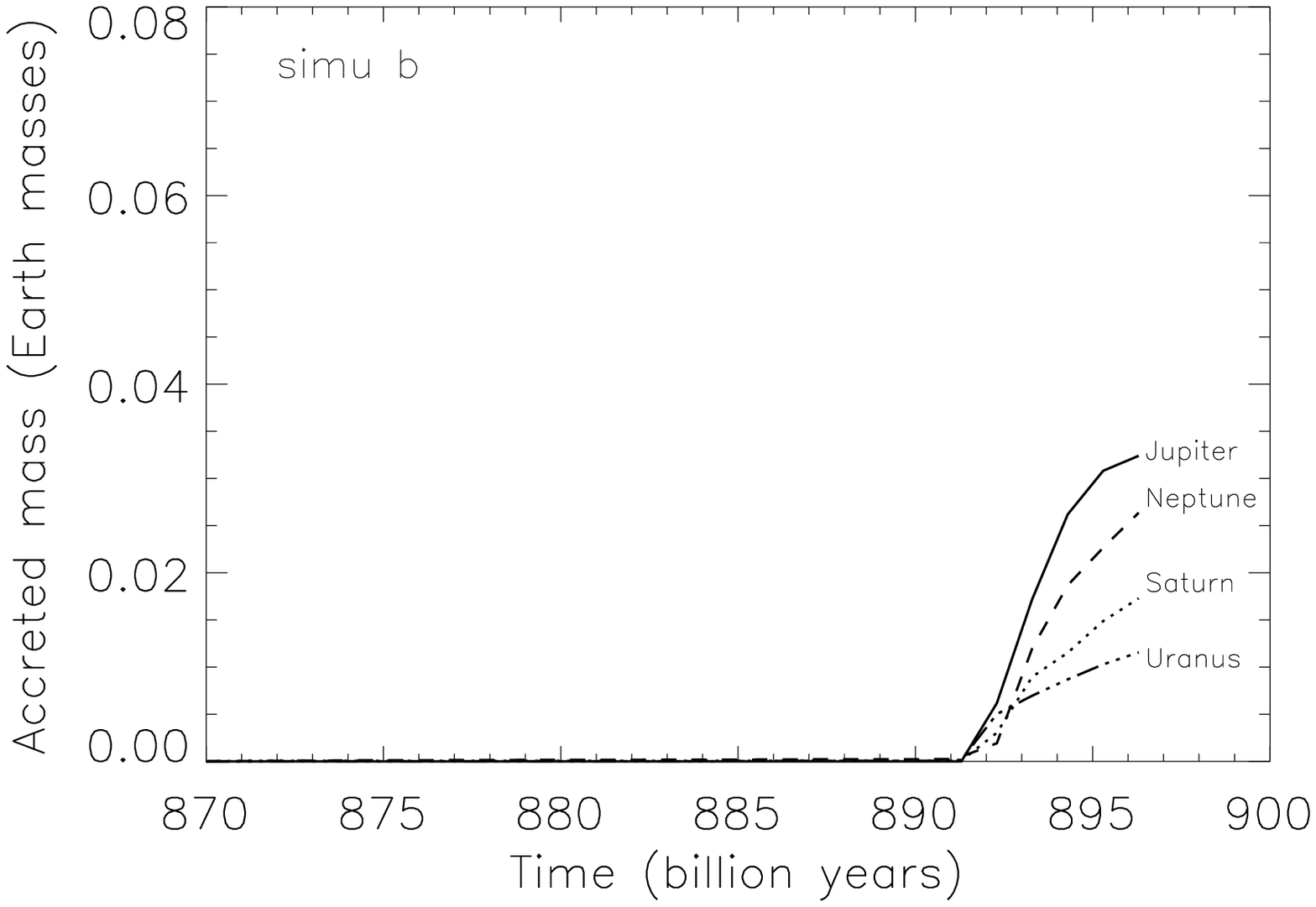}\hfill
\includegraphics[height=55mm,width=70mm,angle=0]{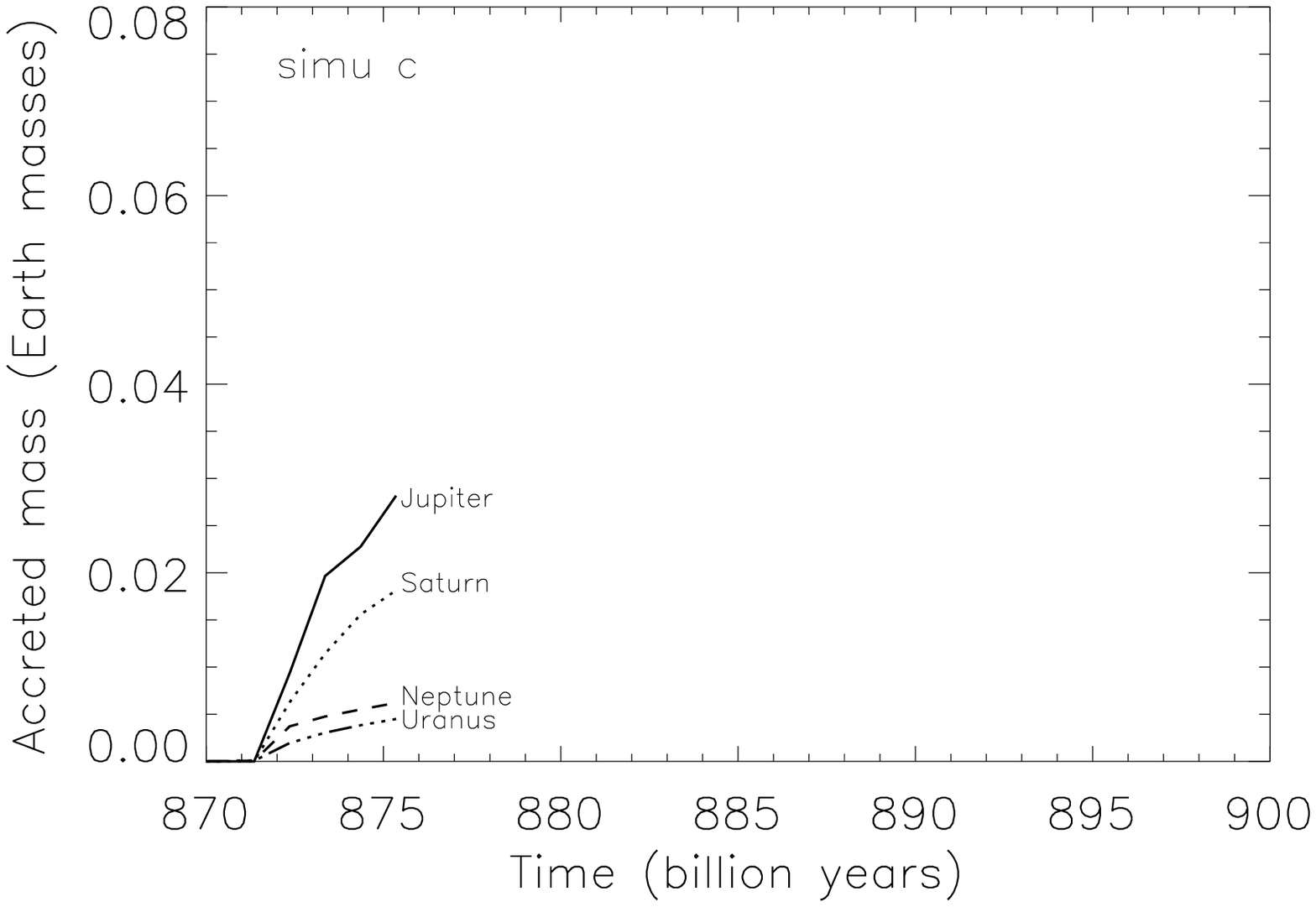}\hfill
\includegraphics[height=55mm,width=70mm,angle=0]{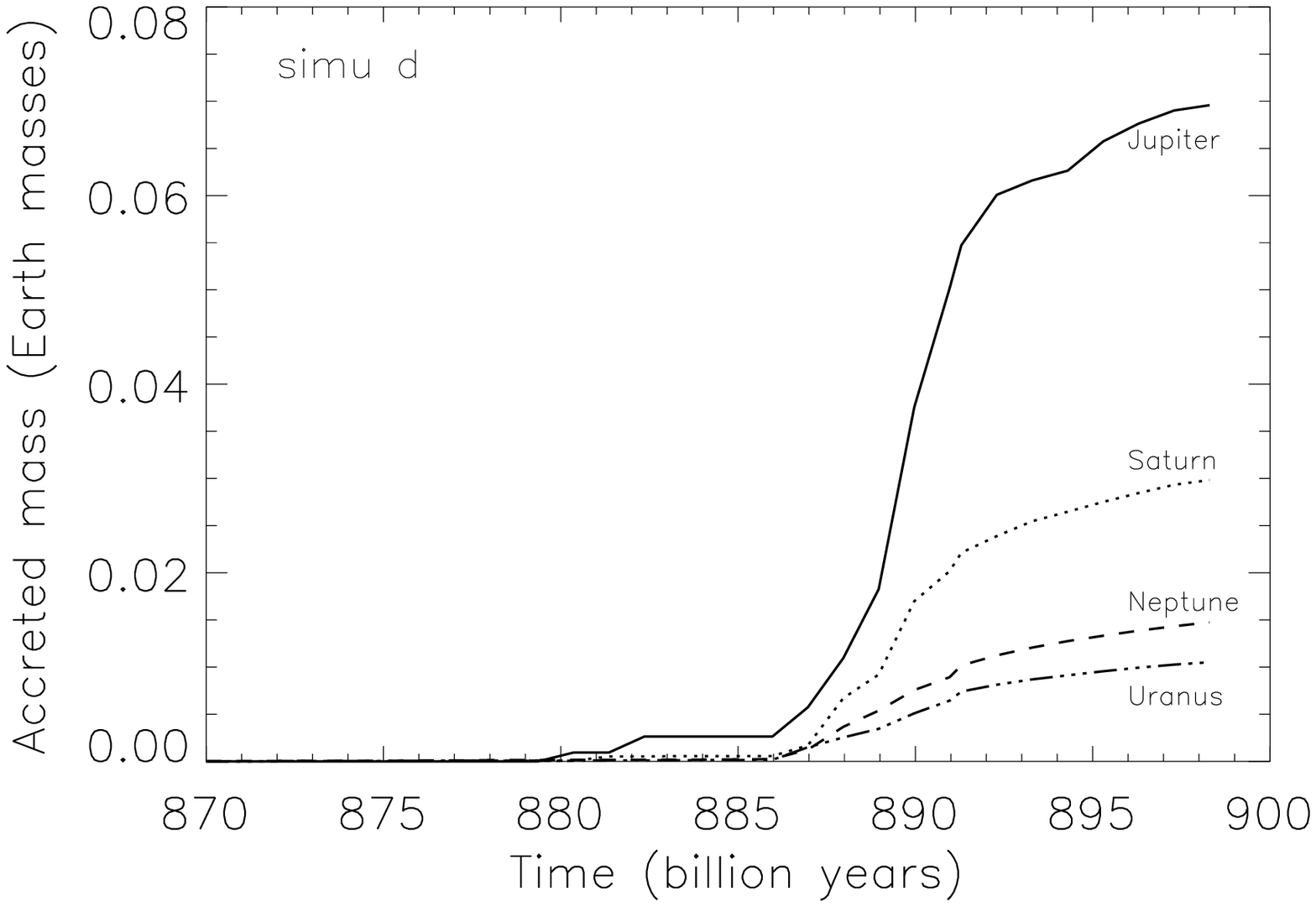}
\label{fig:multiple}
\end{figure} 
  
\begin{figure}[!h]
\caption{}
\includegraphics[height=58mm,width=70mm,angle=0]{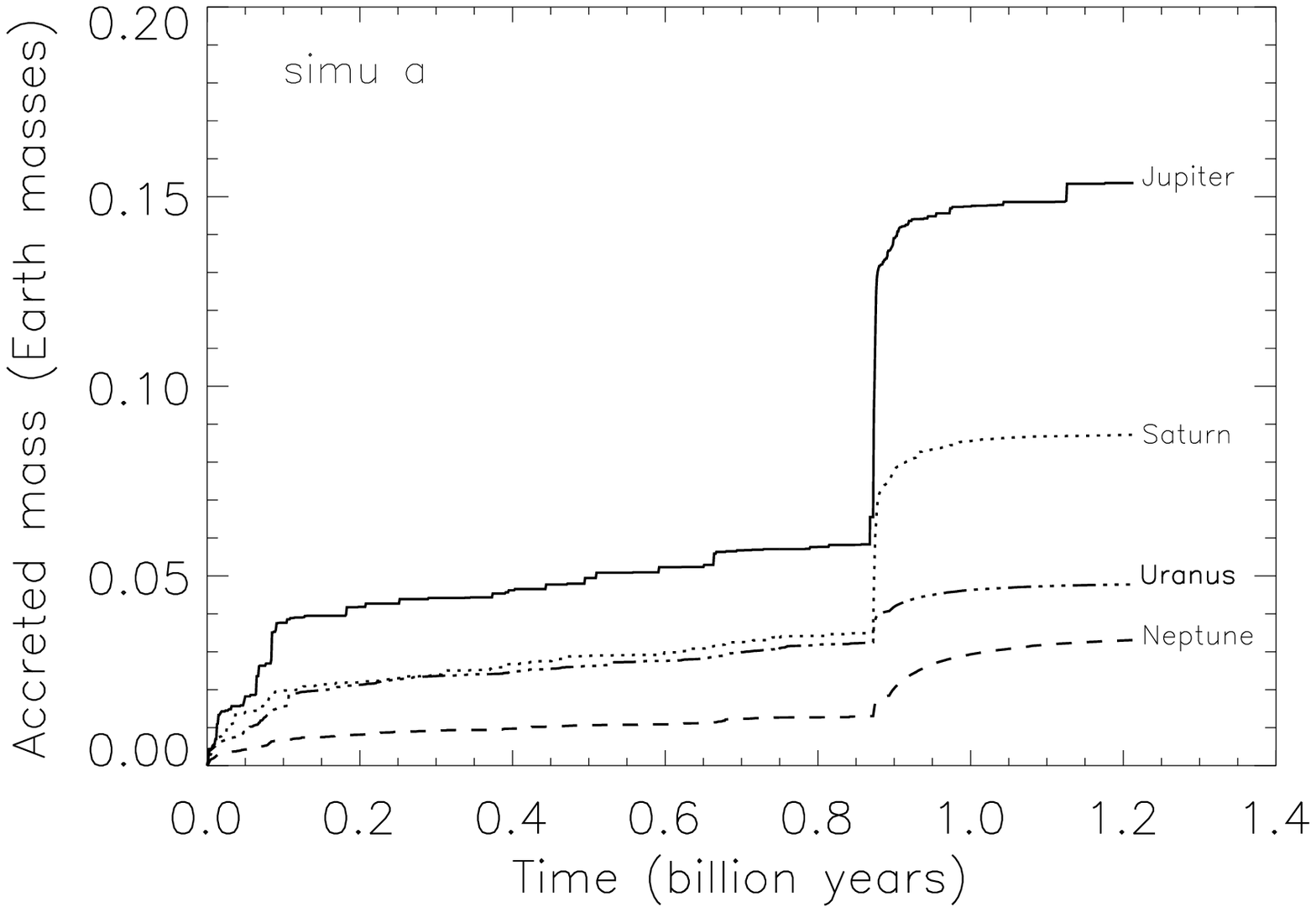}\hfill
\includegraphics[height=58mm,width=70mm,angle=0]{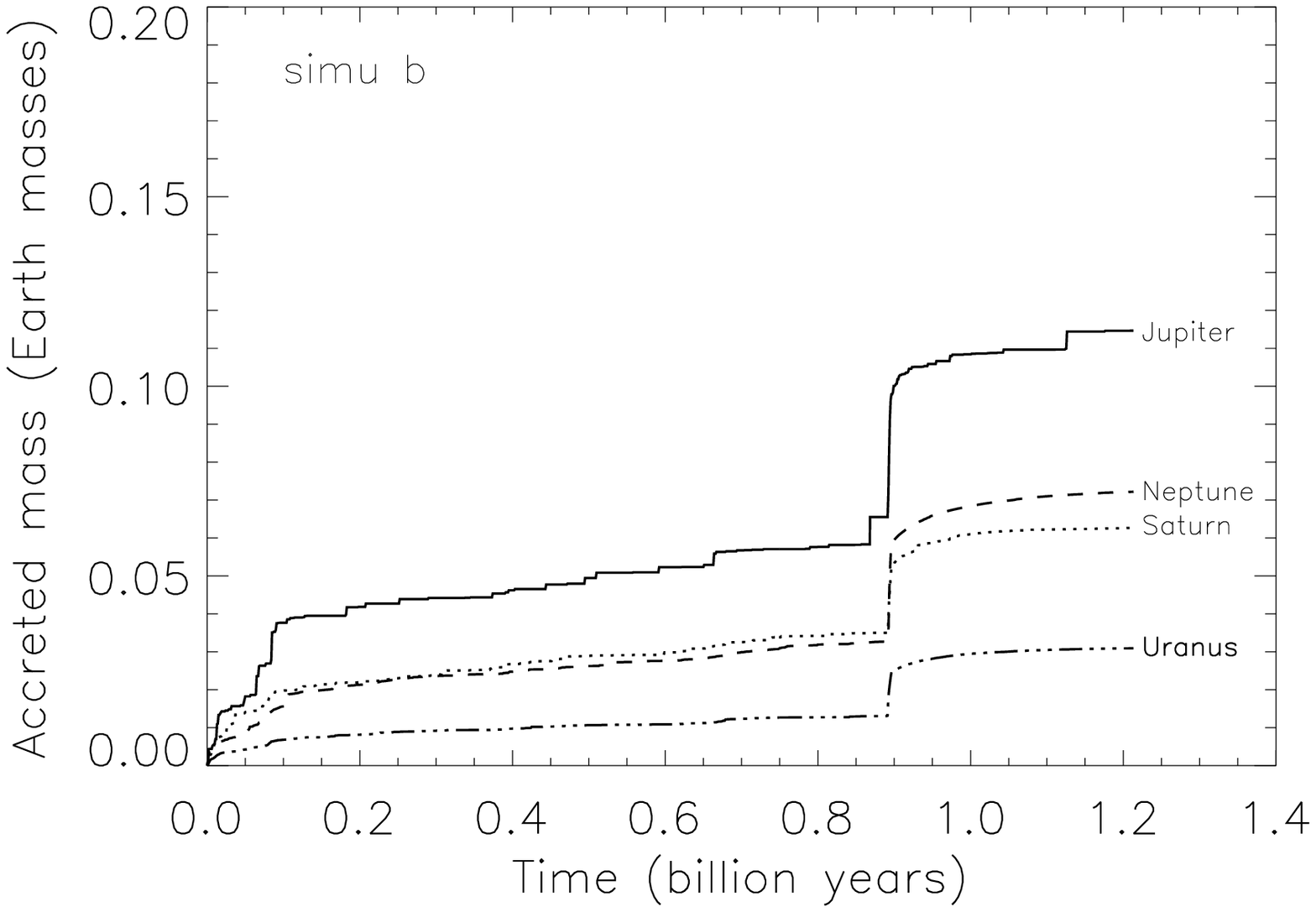}\hfill
\includegraphics[height=55mm,width=70mm,angle=0]{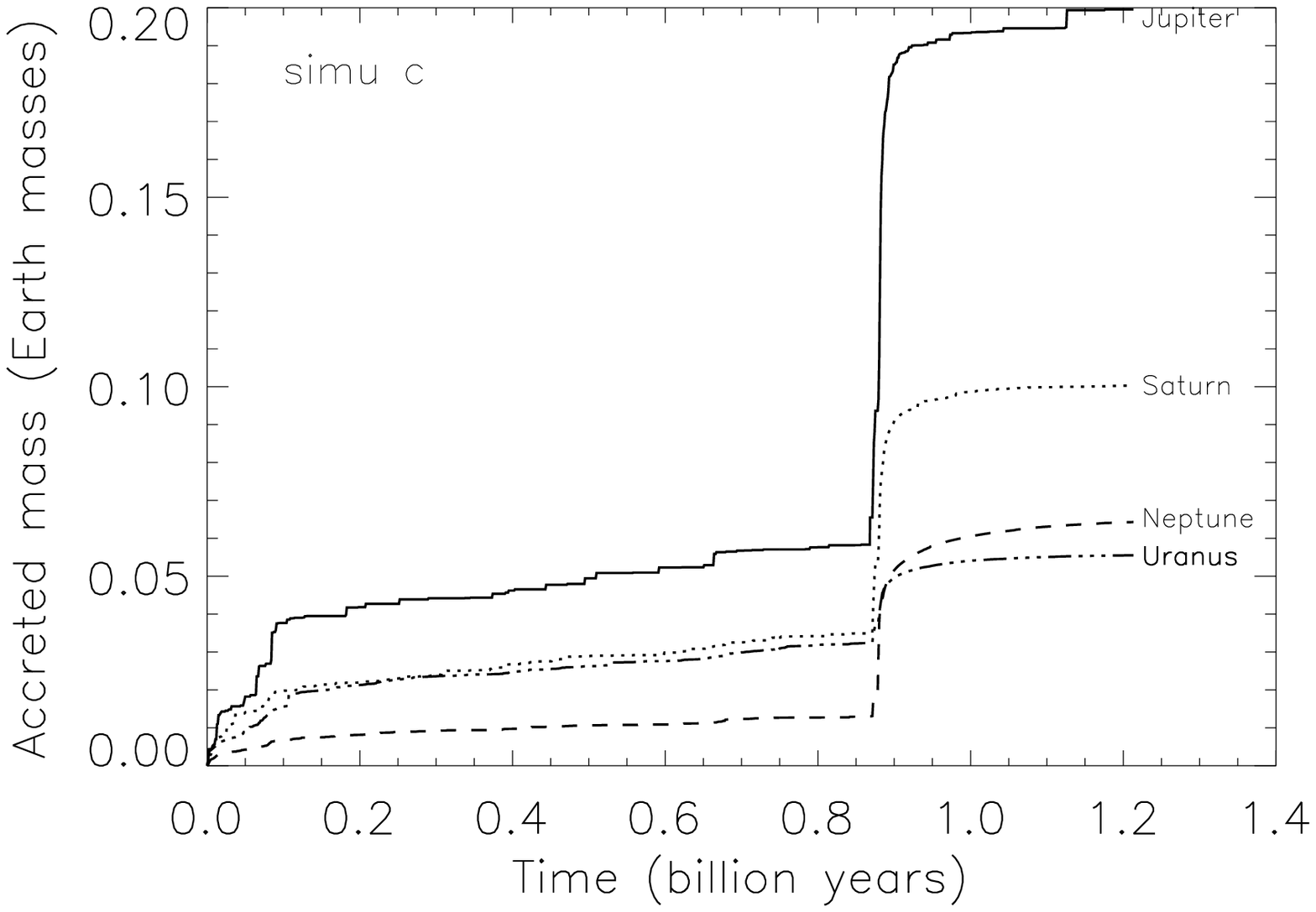}\hfill
\includegraphics[height=55mm,width=70mm,angle=0]{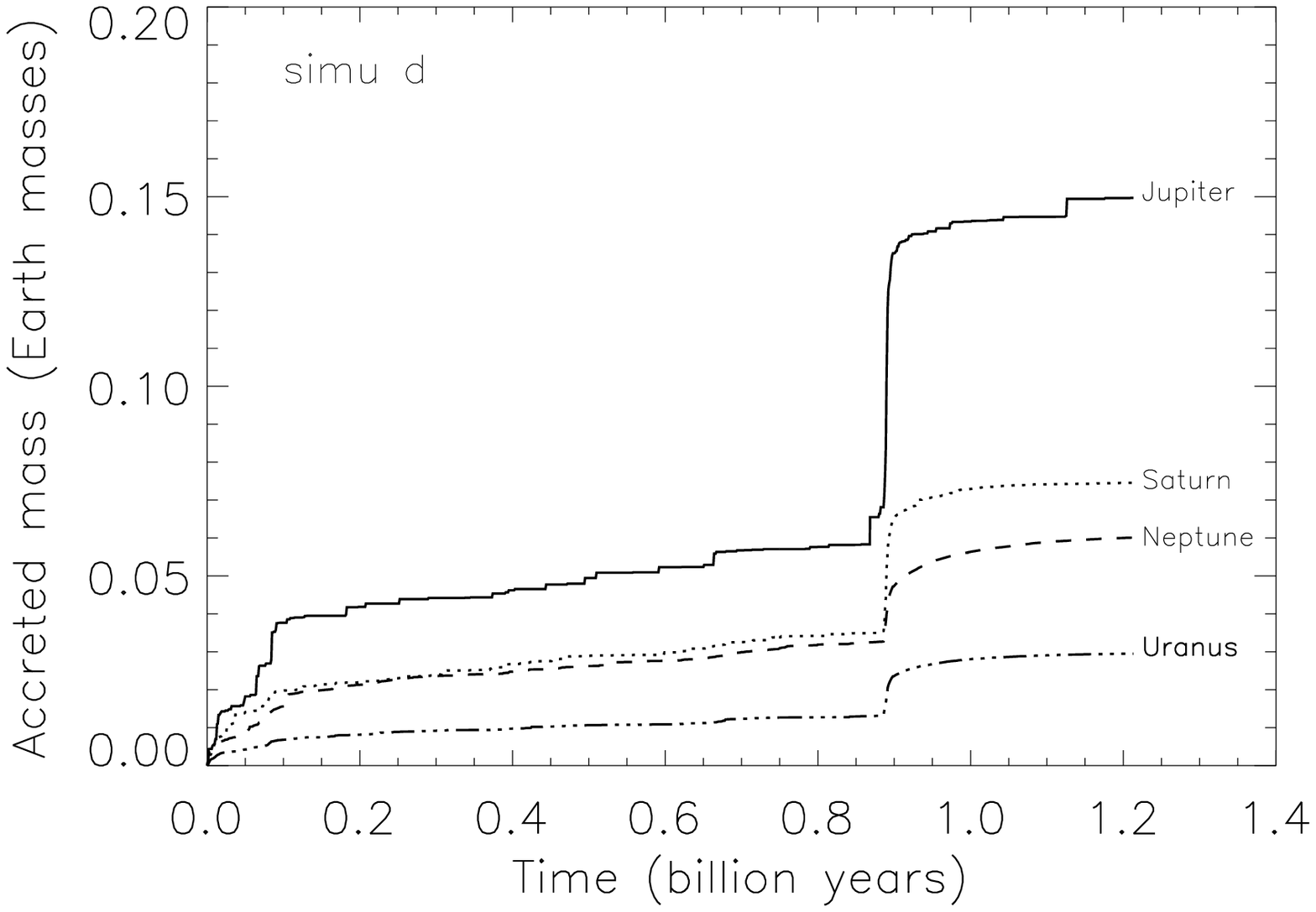}
\label{fig:multiple}
\end{figure} 

\begin{table}
 \begin{center}
 \caption{}
 \label{tab:table}
 \begin{tabular}[!h]{|c|c|c|}
 \hline
 \textit {{Giant planet}}& \textit {{Envelope mass}} $\rm [M_\oplus]$& \textit {{ Accreted mass}} $\rm [M_\oplus]$\\ 
 \hline
  { Jupiter}& { $300-318$}& { $0.11-0.20$} \\
 \hline
  { Saturn}& { $70-85$}& { $0.06-0.10$} \\
 \hline
  { Uranus (w/ inversion)}& { $1-4$}& { $0.048-0.055$} \\
 { \hphantom{Uranus} (w/o inversion)}&   & { $0.029-0.031$} \\
 \hline
  { Neptune (w/ inversion)}& { $1-4$}& {$0.033-0.064$} \\
  {\hphantom{Neptune} (w/o inversion)} &  & {$0.060-0.072$} \\
 \hline
 \end{tabular}
 \label{tab:accreted}
 \end{center}
 \end{table}
 
 \begin{table}
 \centering
 \caption{}
 \begin{tabular}[!h]{|c|c|c|}
 \hline
 \textit {{ Giant planet}}& {$\Delta\varepsilon_{\rm LHB}$}& \textit{Observed $\varepsilon_{\rm C/H}$}\\ 
 \hline
  { Jupiter}& { 0.032-0.034}& 4.1 $\pm$ 1\\
 \hline
  { Saturn}& { 0.062-0.076}& 7.4 $\pm$ 1.7\\
 \hline
 { Uranus (w/ inversion)}& { $0.8-3.4$}& 45 $\pm$ 20 \\
 { \hphantom{Uranus} (w/o inversion)}& { $0.5-2$}& \\
 \hline
  { Neptune (w/ inversion)}& { 0.8-3.2} & 45 $\pm$ 20\\
  {\hphantom{Neptune} (w/o inversion)}& {1.1-4.4}& \\
 \hline
 \end{tabular}
 \label{tab:enrich}
 \end{table} 
 
 \begin{table}
 \centering
 \caption{}
 \begin{tabular}[!h]{|c|c|c|c|c|c|}
 \hline
 \textit {{ Giant planet}}& {\rm M}$_{\rm mixed}$&{\rm h}$_{\rm mixed}$& P$_{\rm mixed}$& s$_{\rm mixed}$& \% $M({\rm s}>{\rm s}_{\rm mixed})$\\
 \hphantom{\textit{Giant planet}}& (M$_\oplus$) & (km) & (bar) &(km) & \\
 \hline
  { Jupiter}& {3.41}& 2000& 48300& 248& 23\%-32\%\\
 \hline
  { Saturn}& {0.82}& {2200}& 7200& 86& 39\%-54\%\\
 \hline
 { Uranus (w/ inversion)}& {0.11}& 1200& 4200& 63& 44\%-60\% \\
 { \hphantom{Uranus} (w/o inversion)}& 0.07& {1000}& 2500& 37& 57\%-69\%\\
 \hline
  { Neptune (w/ inversion)}& {0.11}& 1000 & 5100& 61& 45\%-61\%\\
  {\hphantom{Neptune} (w/o inversion)}& 0.15& 1400 & 13100& 84& 39\%-54\%\\
 \hline
 \end{tabular}
 \label{tab:enrich}
 \end{table}

\end{document}